\documentclass{sig-alternate}
\usepackage{url}
\usepackage{alltt}
\usepackage{subfigure}
\usepackage{float}
\usepackage{amsmath}
\usepackage{algorithm}
\usepackage{algorithmic}
\begin{document}

\title{ Topical Community Detection \\in Event-based Social Network}
\numberofauthors{2}
\author{
\alignauthor
Houda Khrouf\\
       \affaddr{Atos}\\
       \affaddr{Bezons, France}\\
       \email{houda.khrouf@atos.net}
\alignauthor
Rapha\"el Troncy\\
       \affaddr{Eurecom}\\
       \affaddr{Biot, France}\\
       \email{raphael.troncy@eurecom.fr}
}
\maketitle


\begin{abstract}
Event-based services have recently witnessed a rapid growth driving the way people explore and share information of interest. They host a huge amount of users' activities including explicit RSVP, shared photos, comments and social connections. Exploiting these activities to detect communities of similar users is a challenging problem. In reality, a community in event-based social network (ESBN) is a group of users not only sharing common events and friends, but also having similar topical interests. However, such community could not be detected by most of existing methods which mainly draw on link analysis in the network. To address this problem, there is a need to capitalize on the semantics of shared objects along with the structural properties, and to generate overlapping communities rather than disjoint ones. In this paper, we propose to leverage the users' activities around events with the aim to detect communities based on topical clustering and link analysis that maximize a new form of semantic modularity. We particularly highlight the difference between online and offline social interactions, and the influence of event categories to detect communities. Experimental results on real datasets showed that our approach was able to detect semantically meaningful communities compared with existing state of the art methods.
\end{abstract}

\vspace{3mm}
\section{Introduction}
Events are a natural way for referring to any observable occurrence grouping persons, places, times and activities that can be described and documented through different media~\cite{Shaw:ASWC09}. Today's event landscape is increasingly crowded with new websites including event directories, social networks and media platforms. People have been recently attracted by these services to organize and distribute their personal data according to occurring events, to share captured media and to create new social connections. Websites such as Eventful\footnote{\url{http://www.eventful.com}}, Lanyrd\footnote{\url{http://lanyrd.com}}, Last.fm\footnote{\url{http://www.last.fm}}, Flickr and Twitter host an ever increasing amount of event-centric knowledge maintained by rich social interactions. In particular, the event-based social network (ESBN) is different from the traditional social network due to the coexistence of two kinds of social interactions. The former is represented by the typical online activities such as sharing comments, photos and friends. The latter captures the face-to-face social interactions reflecting the physical users' co-participation in events. Typical examples are the academic conferences where researchers interact with other community members with whom they may have common research background~\cite{Li:dasfa11a}. In other words, ESBN is an heterogeneous social network underlying the co-existence of both online and offline social links~\cite{Liu:KDD12}. Meanwhile, the information about these social interactions are spread over multiple websites. For example, people tend to mostly use media platforms (Flickr, Twitter) to share photos and thoughts about events, whereas they express their intent to attend events (RSVP) in online event directories (Eventful, Last.fm). Exploring the overlap between these distributed websites is a key advantage to enhance the social network analysis. 

Community detection is considered as a major topic for analyzing social networks which has recently received a great attention. It aims to uncover the substructures within a network that reveals how individuals interact together and which users are likely to have common interests, occupations and social properties. The information about the underlying communities can be of a great benefit for many tasks such as information diffusion, targeted advertising and collaborative recommendation~\cite{Shaghayegh:RSWEB11}. Broadly speaking, detecting communities is dividing the vertices into groups such that there is a higher density of links within groups than between them~\cite{Clauset:PR04}. To achieve this, most of existing methods focus on network topology and structural properties assuming that the interaction strength of users is the reflection of their proximity/similarity. However, communities detected by those methods often represent users having different interests since no consideration of the topical dimension was made. This problem is accentuated when the users interact with different social objects inducing highly diverse topics in one community. Therefore, there is a need to incorporate the semantic information along with the structural properties for detecting meaningful communities~\cite{Juan:cason11,Zhongying:12}. 

In ESBN, it is ideal to analyze the rich content about users and events in order to discover semantically coherent communities. Moreover, a person is naturally interested in many events which can be associated with multiple topics, so that it is more reasonable to divide users into overlapping groups instead of disjoint ones. Nevertheless, communities produced by topic-driven methods may contain weakly connected users which results in significant loss of social information. An efficient community detection algorithm should therefore cluster individuals who are closely connected and sharing common interested topics. 

In this paper, we propose a novel approach which combines event clustering and link analysis to detect communities. First, we compute event similarity based on social information and content attributes. Then, we use a hierarchical clustering to group events into different topics. A link-based function is defined to determine the effective user attachment to each community. A comparison with existing work shows the efficiency of our algorithm to detect communities optimizing both users connectivity and topical purity. The results also highlight how people interact differently in offline and online ESBN, and how these interactions depend on the event category (e.g conference, concert, etc.).

The rest of the paper is organized as follows. Section \ref{sec:related-work} presents the related work.  We describe our dataset called EventMedia in Section~\ref{sec:eventmedia}. Then, we examine some important properties about ESBN in Section \ref{sec:esbn}.  We describe our algorithm based on event clustering and link analysis in Section \ref{sec:community}. The evaluation of our approach is detailed in Section \ref{sec:experiments}. Finally, we conclude the paper in Section \ref{sec:conclusion}.   

\section{Related Work} \label{sec:related-work}
Community detection has attracted attention in recent years leading to several interesting works. Most of existing works attempt to detect disjoint communities by optimizing different measures and objectives. One popular example is the modularity optimization~\cite{Clauset:PR04,Newman:04} which is used to maximize the connectivity between nodes within one community and minimize the connectivity between groups. Another example is the minimization of a defined cut function in spectral methods~\cite{Scott:05}. These works mostly focus on structural properties and linkage patterns of people and they have been successfully used in some applications. However, they  generate groups of users associated with different semantic topics, making hard the interpretation of users proximity in these communities. 

To overcome the limitation of link-based methods, some studies attempt to exploit topic modeling techniques such as PLSA~\cite{Thomas:99}, LDA~\cite{Blei:MLR03}, AT~\cite{Steyvers:kdd04} that attempt to detect topical communities. For example, the work in~\cite{Li:2013} makes an analogy between the LDA document-topic-word and the user-topic-websites to discover topical communities. The idea behind is that users sharing similar online access pattern tend to belong to same topical group. Resulting clusters are labeled with extracted keywords from websites. This method primarily relies on the link information in a social graph and it is only efficient when regular interaction patterns could be detected. Another approach~\cite{Zhou:WWW06} proposes Community-User-Topic, an extension model of LDA which detects communities using the semantics of content. Communities are represented as random mixtures over users who are associated with a topical distribution. This method does not consider the link information assuming that community members are only sharing common topics. It is evident that both methods could not be applied in real world social network where users' membership is conditioned on their social relationships and their shared interests.

Recently, some works start to investigate the combination of both content and link information. For example, the generative Bayesian model (Topic User Recipient Community Model) presented in~\cite{Sachan:WWW12} combines discussed topics, interaction pattern and network topology to detect topical communities. In~\cite{Zhongying:12},  Zhao et al. proposed to use a modified k-means algorithm (EWKM-Entropy Weighting K-means) to divide social objects into topical clusters. Each cluster contains members involved in associated social objects. A modularity maximization method is then used to detect strongly connected communities in each topical cluster. In this work, we make analogy between these social objects and events, and we extensively compare our algorithm with this approach. 

The last concern in related work is the research on discovering communities in ESBN. Liu et al.~\cite{Liu:KDD12} attempted to resolve the problem of community detection in heterogeneous network. They employed an extended Fiedler method to both consider online and offline social interactions. This method seems efficient to detect cohesive communities, but it is a link-based method where there is no interpretation of detected topics and no consideration of multiple users' memberships. In~\cite{Li:dasfa11a}, the Event-based Community Detection (ECODE)  algorithm tried to enrich the graph with virtual links based on content-based users' similarity. These links aim to enhance connectivity among individuals within same topical community. A hierarchical clustering is then used to group events based on their physical and virtual similarities. Users' memberships are finally determined by their involvement in events of each cluster producing overlapping communities. In the same context, Wang et al.~\cite{Wang:14}, proposed a community detection approach in location-based social network (LSBN). They exploited different features such as user social similarity and venue-user similarity and performed an edge-centric co-clustering which simultaneously discovers overlapping groups of venues and groups of users. To sum up, these different studies provides important insights into detecting communities in ESBN. However, none of them aims to maximize both connectivity strength and topical purity which is the focus of this work.

\section{EventMedia}  \label{sec:eventmedia}
An ever increasing amount of event-centric knowledge is spread over multiple social services, either materialized as calendar of past and upcoming events or illustrated by cross-media items. This opens an opportunity to create an infrastructure unifying event-centric information derived from event directories, media platforms and social networks. EventMedia relies on Semantic Web technologies to create such infrastructure which ensures seamless integration of disparate data sources, some of which overlap in their coverage~\cite{Khrouf:SWJ12}.

EventMedia has been added to the Linked Data\footnote{\url{http://linkeddata.org}} cloud since September 2010. It is obtained from four public event directories (Last.fm, Eventful, Upcoming, Lanyrd) and from two large media directories (Flickr and Twitter). It encapsulates events descriptions associated with media and enriched with background knowledge from external datasets such as DBpedia and Foursquare. 

\begin{figure}[H]
  \centering
  \includegraphics[width=\linewidth]{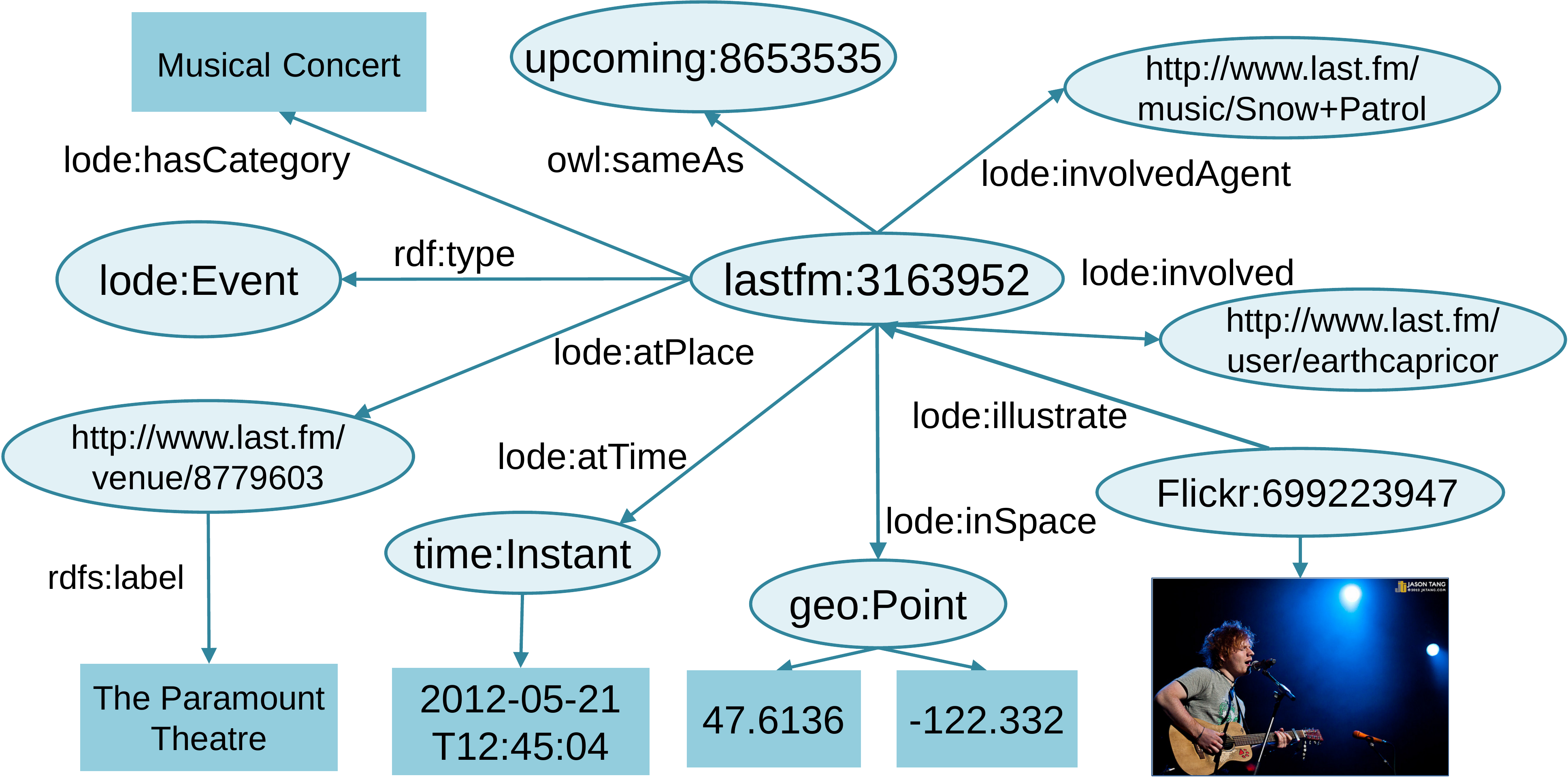}
 \caption{\emph{Snow Patrol Concert} described with LODE ontology}
 \label{fig:lode-ontology}
\end{figure}

In EventMedia, here are more that 30 millions of RDF triples described using some popular vocabularies such as the LODE ontology~\cite{Shaw:ASWC09}, W3C Ontology for Media Resources and Dublin Core. Figure~\ref{fig:lode-ontology} depicts the metadata attached to the event identified by \texttt{3163952} on Last.fm according to the LODE ontology. More precisely, it indicates that an event of type \texttt{Concert} has been given on the \texttt{21th of May 2012 at 12:45 PM} in the \texttt{The Paramount Theatre} featuring the \texttt{Snow Patrol} rock band, and one attendee is the Last.fm user \texttt{earthcapricor}. The connection between events and media is made by the use of existing metadata, namely: (i) the machine tags such as \texttt{lastfm:event=XXX} that connect Flickr with Last.fm and Upcoming; (ii) the hashtags that connect Twitter with Lanyrd repository~\cite{Khrouf:ISWC12,Khrouf:RAMSS12}.

EventMedia contains a highly diverse set of event categories, ranging from large festivals and conferences to small concerts and social gatherings. In this work, we deal with the repositories that provide a considerable number of users, namely the Last.fm and Lanyrd services along with their associated media sites. On one hand, Last.fm is the oldest and largest music based social networking site which is created in 2002. Users can add new musical events which will be listed on the band or artist's page along with other valuable details such as event description, location, tags, etc. On the other hand, Lanyrd\footnote{\url{http://lanyrd.com}} exposes information about past and upcoming conferences ranging from large events such as TED\footnote{\url{http://www.ted.com/}} to smaller ones. Table \ref{tab:dataset-tab} shows some statistics about collected data from Last.fm and Lanyrd repositories as well as their associated media services in EventMedia. We note that EventMedia contains the Last.fm events which are only associated with media from Flickr, and the conferences from Lanyrd happened between February and August 2012.

\begin{table}[H]
\centering{
\begin{tabular*}{8.2cm}{l| @{\extracolsep{\fill}}rccc}
\hline
           & Event & Event User & Media & Media User\\
\hline
   Last.fm  &  {   } 66,757 & 180,673 & 1,530,895 & 20,030 \\
\hline
   Lanyrd & 2,151 & - & 1,030,770 & 261,867 \\
\hline
\end{tabular*}
\caption{Number of resources per type in Last.fm and Lanyrd sub-directories in EventMedia}
\label{tab:dataset-tab}
}
\end{table}

\section{Event-based Social Network} \label{sec:esbn}
In this section, we describe how to construct an event-based network using offline and online interactions (Section \ref{sec:esbn-def}) and we highlight some of their interesting properties (Sections \ref{sec:spatial-aspect} and \ref{sec:network-prop}).

\subsection{ESBN Definition}   \label{sec:esbn-def}

Based on users' activities in social services, we define the following ESBNs making difference between online and offline networks. Slightly different from the definition described by Liu et al.~\cite{Liu:KDD12}, we consider that the online ESBN is constructed by solely capturing the online interactions such as sharing comments and photos about events. This online ESBN is different from the online ``friendship'' social network that may exist in some services. Similarly, the offline ESBN is constructed by considering the physical co-participation in social events.

\vspace{2mm}

{\textbf{Last.fm ESBN.}} In Last.fm, the online ESBN is built from the online co-commenting of social events, whereas the offline ESBN is based on the explict RSVP provided by the users. Besides to these both ESBN, we also consider the social undirected network of friends for comparison purposes.

\vspace{2mm}
{\textbf{Flickr ESBN.}} Flickr is one of the most important online photo and video sharing websites. We leverage the activity of co-sharing photos about events to build an online ESBN. 

\vspace{2mm}
{\textbf{Twitter (Lanyrd) ESBN}} Twitter is a popular micro-blogging service, and it is by far the most used back-channel for commenting scientific conferences~\cite{Khrouf:RAMSS12}. In a similar way, we exploit the co-commenting of conferences to construct an online ESBN.

\subsection{Spatial Aspect of Social Interactions}   \label{sec:spatial-aspect}
\begin{figure}[H]
  \centering
  \includegraphics[scale=0.35]{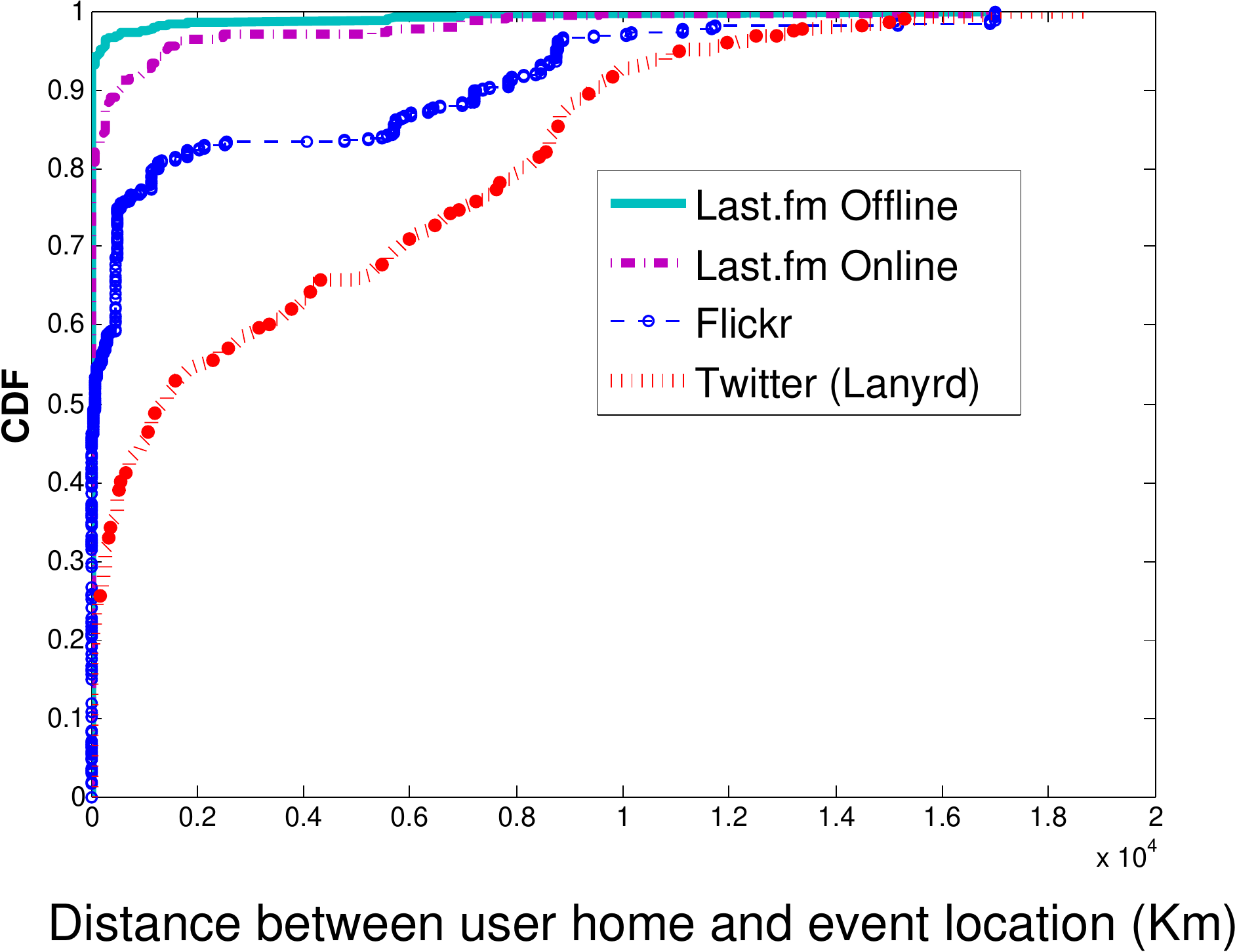}
  \caption{Locality of users' interactions in offline and online ESBN}
  \label{fig:event-user-distance}
\end{figure}

In the following, we investigate how far from their homes people interact in ESBNs. Therefore, we compare the geographical distance between an event location and the user's home in offline and online networks. As the user home location is not explicitly provided in Last.fm, we infer it using the average of most frequent positions of attended events. We show the results in Figure~\ref{fig:event-user-distance} based on a random set of events and their associated users.

 We observe that 95\% of users' activities in offline network are within 100 km. This rate slightly decreases in online Last.fm ESBN indicating that people tend to comment nearby events. This aspect has already been proved in an existing study~\cite{Liu:KDD12} showing that users' activities in ESBNs are much more location constrained compared with location-based social network. In contrast, the online interactions in media-based ESBNs seem to be less conditioned on event location. This can be explained by two reasons: (i) the nature of sharing (retweeting) activity where users are non-uniformly spread; (ii) the type of events indicating that people tend to travel far from their home for business purpose (conference) rather than for entertainment activity (musical concert).  

Based on these findings, we decided to perform community detection using conferences from different cities in Lanyrd, whereas we only focus on a specific geographical location in Last.fm.


\subsection{User Participation}   \label{sec:network-prop}

\begin{figure}[H]
  \centering
  \includegraphics[scale=0.45]{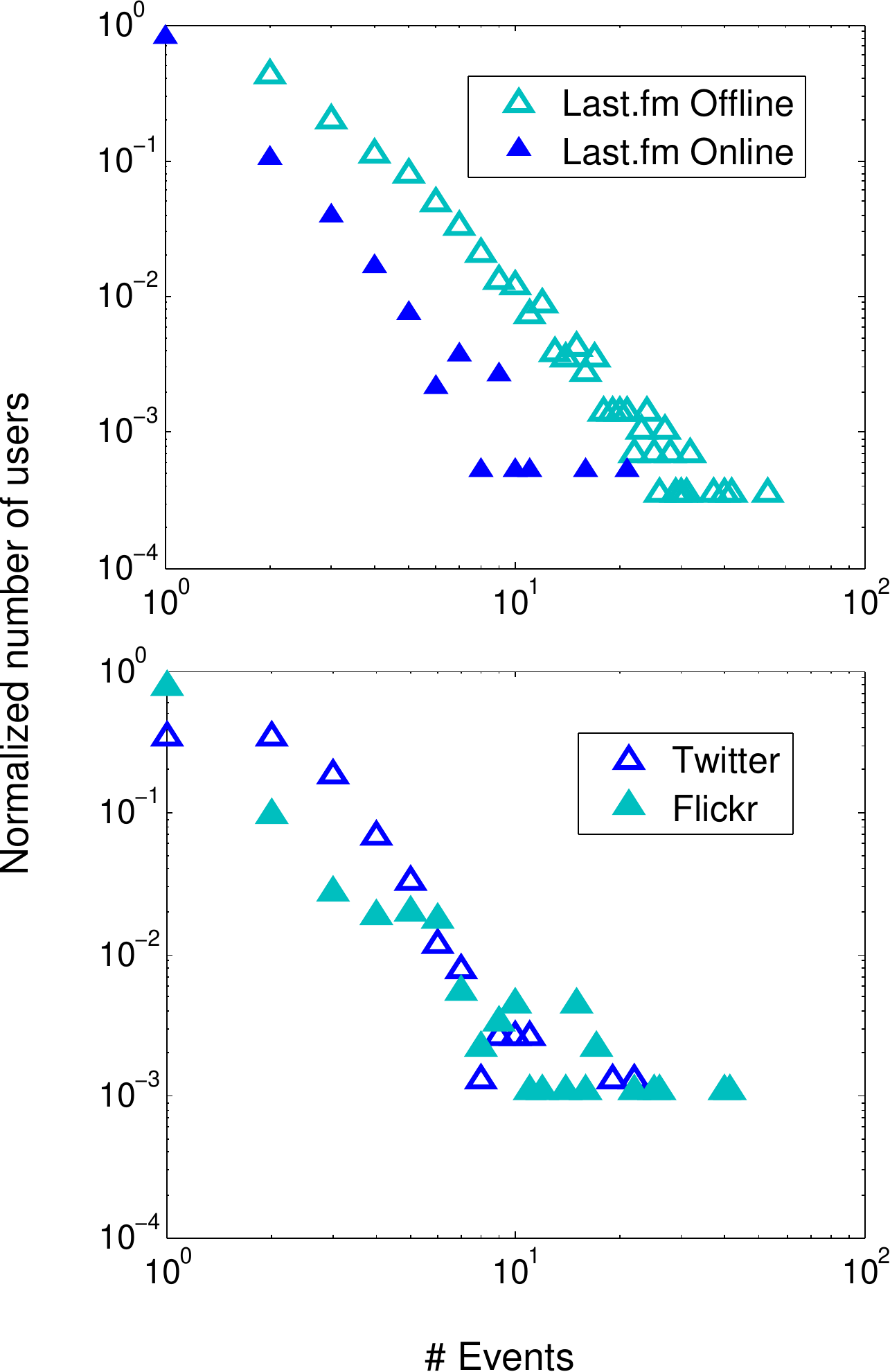}
  \caption{Number of participants per event}
  \label{fig:users-events}
\end{figure}

To gain insights into some properties of the described ESBNs, we study the user participation behavior. As shown in Figure~\ref{fig:users-events}, the results resemble a power-law distribution indicating that most of users are associated with few events. Similar results have been highlighted in other works about event attendance behavior~\cite{Liu:KDD12,Han:NCWTW12}. In particular, there are 81\% of users who are associated with only one event in Last.fm online ESBN, and 76\% of users sharing photos of only one event in Flickr ESBN (this is also proved in Table~\ref{tab:dataset-tab} when we compare the number of media shared and their associated users). During the evaluation (section~\ref{sec:experiments}), we will show the impact of user participation behavioir on community detection.    


\section{Topical Community Detection} \label{sec:community}
In this section, we firstly describe our graph model (Section~\ref{sec:modeling}). Then, we present our approach proposed for topical community detection based on three steps: similarity computation, event clustering and users' assignment to communities (Section~\ref{sec:approach}). 

\subsection{Graph Modeling}  \label{sec:modeling}
Taking into account the users, events and related attributes, we consider the three-tuple graph $G=<U,S,T,E>$ for both online and offline ESBN where $U$ is the set of users, $S$ is the set of social events which are in turn associated with a set of tags, and finally $E$ is the set of undirected edges. $E$ contains two kinds of links  $E=E_{US} \cup E_{UU}$ where $E_{US}$ denote the links between users and events representing the user participation in social events $E_{US}={\lbrace(u,s)| u \in U,s \in S \rbrace}$, and $E_{UU}$ is the set of links between users obtained from the co-participation in the same social events where $E_{UU}={\lbrace(u_{i},u_{j})| u_{i} \in U,u_{j} \in U \rbrace}$. 

In this graph, each user can be represented as a vector of events, and each event can be represented as a vector of users. Similar way is applied using the event-tag relationship. We exploit these representations to compute the similarity of events which will be used for detecting communities.

\subsection{The Proposed Approach}   \label{sec:approach}
In our approach, we follow the same logic as EWKM-based method proposed in~\cite{Zhongying:12}. The idea behind this method is to firstly cluster the social objects from topical perspective, and then it clusters associated users into groups having higher modularity. Rather than using two-step clustering, we propose one step clustering taking into account both the link and content attributes. 

\subsubsection{Similarity Computation}  \label{sec:similarity}
In ESBN, overlapping communities of users who share same interests can be detected by clustering similar events together~\cite{Li:dasfa11a}. Moreover, considering the number of events and users, we assume that event-based clustering have a less computational time compared with user-based clustering. To discover topical communities, the event similarity should reflect both the link and content information. In this context, we use the notion of \emph{Homophily}~\cite{McPherson:2001} observed in many social networks: the users involved in same events have a higher likelihood to get connected. Similarly, the tags associated with same events are more likely to be topically similar. This implies that similar events are sharing both like-minded users and semantically similar tags. 

In an event-user network, events can be represented as a vector of users, and users can also be viewed as a vector of events. To reduce the dimension of the event-user matrix, we need to represent events in a latent user space using an orthogonal basis. Singular Value Decomposition (SVD) is one popular technique employed to obtain such basis.  Given a matrix $A$, the singular value decomposition is the product  $U \Sigma V^T$ where $U$ and $V$ are the left and right singular vectors and $\Sigma$ is the diagonal matrix of singular values. Event vectors in the latent user space is represented by the matrix $\tilde{A}$ illustrated in Equation~\ref{eq:event-latent}.

\begin{equation}  \label{eq:event-latent}
\tilde{A}=U\Sigma \Rightarrow  \tilde{A}=AV
\end{equation}

To detect similar events sharing like-minded users, we leverage the spectral co-clustering~\cite{Dhillon:KDD01} indicating that only the top singular vectors except of the principle one contain partition information. The algorithm first normalizes the event-user matrix: 

\begin{equation} 
A_n=D_1^{-1/2} A D_2^{-1/2} 
\end{equation}

where the entries of the diagonal matrices $D_{1} $ and $D_{2}$ are respectively the event degrees and user degrees. Then, applying singular value decomposition  gives $A_n=U_n \Sigma_n V_n^T$.  Only the top-k singular vectors (except of the principle one) are selected from $V_n$ to form $V'_n$ matrix. Finally, the event representation in user latent space is shown in Equation~\ref{eq:event-latent-1}.

\begin{equation}  \label{eq:event-latent-1}
\tilde{A}=D_1^{-1/2} A_n V'_n
\end{equation}

Applying this algorithm on event-tag network, we have been able to also represent events in a latent semantic space. Then, the similarity of events $S_u$ and $S_t$ respectively in the latent user and semantic spaces are computed using Cosine distance. Finally, we combine the similarities as shown in Equation~\ref{eq:event-sim}.

\begin{equation}  \label{eq:event-sim}
S_{sim} =\alpha \ S_u + (1-\alpha) \ S_t
\end{equation}

In this approach, pair-wise similarities could be reduced by selecting candidate relevant set which index the potentially similar events. This set is represented by events that share in common a minimum number of tags or users. In~\cite{Li:dasfa11a}, it has been shown that this method was efficient to save a significant amount of computational time and could be easily applied in large networks. However, we do not adopt the candidate selection in this work as we deal with small datasets.

\subsubsection{Hierarchical Clustering} \label{sec:clustering}
Inspired by ECODE algorithm~\cite{Li:dasfa11a}, we use a hierarchical agglomerative clustering to group similar events in terms of ``correlated'' users and tags. As outlined in Algorithm~\ref{alg:clustering}, the most similar events $s_i$ and $s_j$ are clustered together forming a new event $s_{new}$. Then, we compute the similarities between $s_{new}$ and each event in the dataset or in the candidate set (if the candidate selection is adopted). The clustering stops when there is no significant increase of the quality function. This approach is advantageous compared with other algorithm such as k-means since the predefined number of clusters is not required.

To produce the minimum number of topics in each cluster, the quality function of the tree follows the same rationale than Newman modularity, but applied in semantic space instead of link-based space. Thus, we aim to maximize the intra-similarities and minimize the inter-similarities in the semantic space. We leverage the event similarity  $S_t$ computed in the latent semantic space and we compute the intra-similarities (Equation~\ref{eq:intra}) and inter-similarities (Equation~\ref{eq:inter})  as following:

\begin{equation} \label{eq:intra}
       {IntraSem}=\frac{1}{|C|}\sum_{C_k \in C} \frac{\sum_{i,j \in C_k, i  C_i j } S_t(i,j)}{|C_k| (|C_k|-1)} 
\end{equation}

 \begin{equation} \label{eq:inter}
      {InterSem}=  \frac{1}{|C|} \sum_{i \in C_i} \frac{\sum_{j \in C_j,  C_i \ne  C_j} S_t(i,j)^2}{M}  
\end{equation}

where C is the set of discovered clusters and M is the number of comparisons in inter-similarity. Finally, we formalize the new semantic modularity SemQ in Equation~\ref{eq:quality}.

 \begin{equation} \label{eq:quality}
		SemQ=IntraSem-InterSem
\end{equation}

\begin{algorithm}
\caption{Agglomerative clustering of similar events}
\label{alg:clustering}
\begin{algorithmic}
\STATE S: set of social events $s_1, s_2...s_i$
\STATE T: number of topics
\STATE $S_{sim}$: event similarity matrix
\WHILE {Community Size>T \AND  SemQ function increases }
	    \STATE Merge the most similar events $s_i$ and $s_j$ into a new event $s_{new}$
 		\FOR { each event $s_k$ $\in$ S (or candidate set)}
 			\STATE $S_{sim}(s_{new},s_k)$ =  average($S_{sim}(s_{new},s_i)$,$S_{sim}(s_{new},s_j)$)
		\ENDFOR
\STATE Compute SemQ function		
\ENDWHILE
\end{algorithmic}
\end{algorithm}

Note that the maximal SemQ measure will provide topical clusters of events, which stops the clustering process. Moreover, each detected cluster keeps in mind a minimal knowledge about the link information and the content information, which makes our approach different from EWKM-based~\cite{Zhongying:12} and ECODE~\cite{Li:dasfa11a} algorithms. 

\subsubsection{User Assignment}   \label{sec:assignment}
The last step of our approach is to group together the participants involved in each event cluster. As the user may participate in many events, we generate overlapping topical communities. However, a user can be weakly involved in one topical cluster which not really reflects his interests. To address this problem, we propose to discover the effective users' membership by computing the assignment scores. If the user $u_i$ is a member of the community $C_i$, the assignment score is defined as follows:

\begin{equation}    \label{eq:assignment}
			AS(u_i,C_i)=\frac{D_c(u_i)}{D(u_i)}				
\end{equation}

where $D_c(u_i)$ is the degree of the user $u_i$ within the community $C_i$, and $D(u_i)$ is the global $u_i$'s degree. The user membership to one community is determined if the assignment score is higher than the average of non-zeros scores over all communities. Note that the user assignment based on Equation~\ref{eq:assignment} may convert a cluster to an empty one. We believe that the removal of these empty clusters is reasonable since they represent a group of very weakly connected users.

\section{Experiments and Evaluation}   \label{sec:experiments}
This section presents the evaluation of the proposed community detection approach by performing experiments on real datasets. We first describe the data collection, followed by the introduction of the performance metrics and the obtained results.

\subsection{Experimental Datasets}
Based on the definitions of the online and offline ESBNs, we use the following datasets\footnote{\url{http://www.eurecom.fr/~khrouf/esbn}} (some statistics shown in Table~\ref{tab:stats}).

\begin{table}[H]
\centering{
\begin{tabular}{cccc}
\hline
& Edges & Density & ClustCoeff \\
\hline
Last.fm Offline & 95897 & 0.0237 & 0.1144\\
\hline
Last.fm Online & 9936 & 0.0067 & 0.398\\
\hline
Flickr Online &  7071 & 0.0188 & 0.2624 \\
\hline
Lanyrd Online & 14237 & 0.0483  & 0.4852 \\
\hline
\end{tabular}
\caption{Some datasets statistics}
\label{tab:stats}
}
\end{table}
\textbf{Entertainment (Last.fm and Flickr):} We previously demonstrate that a very high fraction of social interactions for entertainment purpose  exist between geographically close friends. Hence, we focus our analysis on events located in one city, and we select the capital of England ``London'' as it exhibits a significant number of users and events compared with other cities in EventMedia. We retrieved data using SPARQL queries on EventMedia endpoint\footnote{\url{http://eventmedia.eurecom.fr/sparql}}, and we also crawled additional metadata using the REST API of Last.fm and Flickr. Then, we pre-processed the dataset as follows: First, we removed the tags with very low frequency (less than 5) to reduce the topical noise, and we only keep events which are associated with frequent tags (musical genres). Second, we removed the singletons of event-user pairs where the event has only one participant, and this participant is associated with only one event. We retrieved the events happened in 2012 and 2013 (associated with media) and we obtained the following ESBNs: (i) an offline Last.fm ESBN containing 915 events, 2847 users and 272 tags; (ii) The associated online Last.fm ESBN contains 470 events (among 915 events), 1729 users and 248 tags (among 272 tags); (ii) The associated online Flickr ESBN contains 375 events, 868 users and 221 tags. Note that the removal of the singletons event-user pairs has significantly reduced the size of the online Last.fm and Flickr ESBNs indicating that online users' activities are more sporadic and represent larger individual behaviors than the offline activities.

\textbf{Conference (Lanyrd and Twitter)} In a similar way, we used SPARQL queries and Twitter API to retrieve data. Note that Lanyrd also provide details about the conference attendees, but this information was missing in EventMedia at the time of writing. Thus, we plan to further enrich our dataset and we left the analysis of offline Lanyrd ESBN for future study. We pre-processed the data retrieved as follows. As no tags are associated with events, we automatically processed the conference description (tokenization, filtering, etc.). However, very noisy tags have been produced as some conferences are vaguely described (e.g \emph{The World is Changing, Is Your Company on Board?}).  Similarly, the automatic processing of tweets generate many tags which do not really reflect what the conference about. To overcome this problem, we manually label the conference descriptions selecting the most representative keywords. As there is a manual effort, we tried to only keep the interesting conferences which are related with very active users. Finally, the online ESBN retrieved contains 275 events, 768 users and 166 tags. Note that we obtained a small set of events compared with Last.fm ESBNs due to the high selectivity followed.

\subsection{Topic Modeling}
In order to evaluate the topical purity of each cluster, we need first to detect the set of topics in each dataset. To do this, we decided to employ LDA~\cite{Blei:MLR03}, a popular topic modeling technique where we consider the events as documents. The use of LDA has led to coherent topics in Lanyrd dataset, but noisy and ambiguous in Last.fm datasets. The reason behind these results lies in the nature of events considered and the manual labeling in Lanyrd dataset. Indeed, LDA is sensitive to the co-occurrence of terms in the documents, which results in confused distribution when the documents are topically diverse. In Last.fm, events are musical concerts which feature artists that may share different genres (i.e topics) or only one genre. Moreover, a broad musical genre have different sub-genres that are ``topically'' close but having different labels. In contrast, the conferences are different from musical concerts as they target general one main topic. To solve the topic modeling in Last.fm dataset, we decided to exploit the existing SKOS\footnote{\url{http://www.w3.org/2004/02/skos/}} taxonomy of musical genres in DBpedia using the generalization relations between genres (e.g. \texttt{skos:narrower/skos:broader}).  

\begin{table}[H]
\centering{
\begin{tabular}{c|c}
\hline
    \textbf{Topic}       &  \textbf{Example of Lanyrd Tags}  \\
\hline
   Education  &   learning, education, teaching, technology\\
\hline
   programming  &   programming, language, python, library\\
\hline
   Innovation  &   creativity, technology, business, future\\
\hline
   Application  &   mobile, application, web \\
\hline
\end{tabular}
\caption{Example of topics detected in Lanyrd}
\label{tab:topics-lanyrd}
}
\end{table}

\begin{table}[H]
\centering{
\begin{tabular}{c|c}
\hline
    \textbf{Topic}       &  \textbf{Example of Last.fm Tags}  \\
\hline
   Heavy metal  &   metal alternative, progressive metal... \\
\hline
   Pop  &   synthpop, powerpop, pop punk...\\
\hline
   Electronic  &   indietronica, synthpop, folktronica...\\
\hline
   Rock  &    hard rock, alternative rock, glam rock... \\
\hline
\end{tabular}
\caption{Example of topics detected in Last.fm}
\label{tab:topics-lfm}
}
\end{table}

\begin{figure}[H]
  \centering
  \includegraphics[scale=0.42]{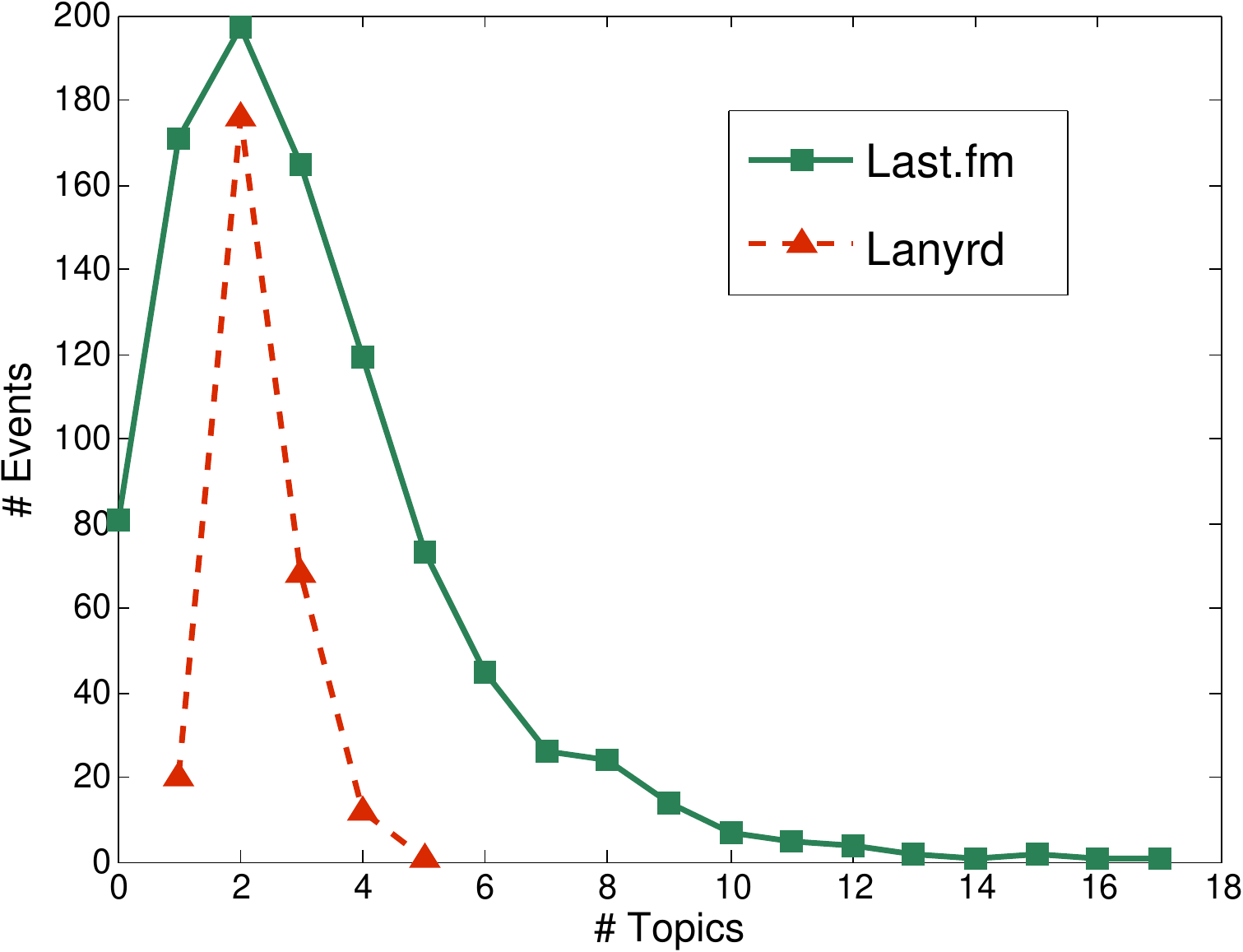}
  \caption{Histogram of the number of topics per event}
  \label{fig:event-topic}
\end{figure}

Tables \ref{tab:topics-lanyrd} and \ref{tab:topics-lfm} show few examples of topics detected respectively in Lanyrd and Last.fm. Note that we obtained 24 topics in Last.fm consisting of high-level musical genres, and 30 topics in Lanyrd where the optimal number of topics is determined based on Griffiths et al. approach~\cite{Griffiths:04}. Finally, we observe in Figure~\ref{fig:event-topic} that most of conferences have at most two topics in Lanyrd, while a slightly higher topical diversity exists in Last.fm events. 

\subsection{Evaluation Metrics}
To evaluate our approach, the performance metric should consider the combination of both content and link information. We adopted the $PurQ_\beta$ metric defined in~\cite{Zhongying:12}. It is inspired by F-score measure which considers both the precision and the recall widely used in information retrieval. Similarly, $PurQ_\beta$ attempts to consider both the topical purity and the members connectivity. Hence, we first define the topical Purity of each cluster as following:

\begin{equation}    \label{eq:purity}
			Purity_i= max_j \left( \frac{n_{ij}}{n_i} \right)
\end{equation}

where $n_{ij}$ is the number of tags belonging to topic $j$ and cluster $i$, and $n_i$ is the number of tags in cluster $i$. The final score of $Purity$ is the average purity scores of all clusters. Yet, we observe during the experiments that the Purity measure does not effectively reflect the presence of clusters having low topical purity. Hence, we decided to also examine the $F_{purity}$ which is the faction of clusters having $Purity_i$ higher or equal than the average $Purity$. Finally, $PurQ_\beta$ is illustrated in Equation \ref{eq:purq}.

\begin{equation}    \label{eq:purq}
			PurQ_\beta= \frac{(1+\beta^2) (Purity \times Q )}{\beta^2 Purity + Q}			
\end{equation}

where $Q$ is the Newman modularity~\cite{Newman:04} used to evaluate the goodness of a partition, ensuring that there are many edges within communities and only a few between them. Then, the parameter $\beta$ is used to adjust the weight of $Purity$ and $Q$. $\beta=0.5$ means that $PurQ_\beta$ puts more emphasis on $Purity$ than $Q$. In contrast, $\beta=2$ puts more emphasis on $Q$. The general behavior of communities is when $Purity$ increases, $Q$ decreases and vice versa.

\subsection{Results} 
\begin{figure}[H]
  \centering
  \includegraphics[scale=0.45]{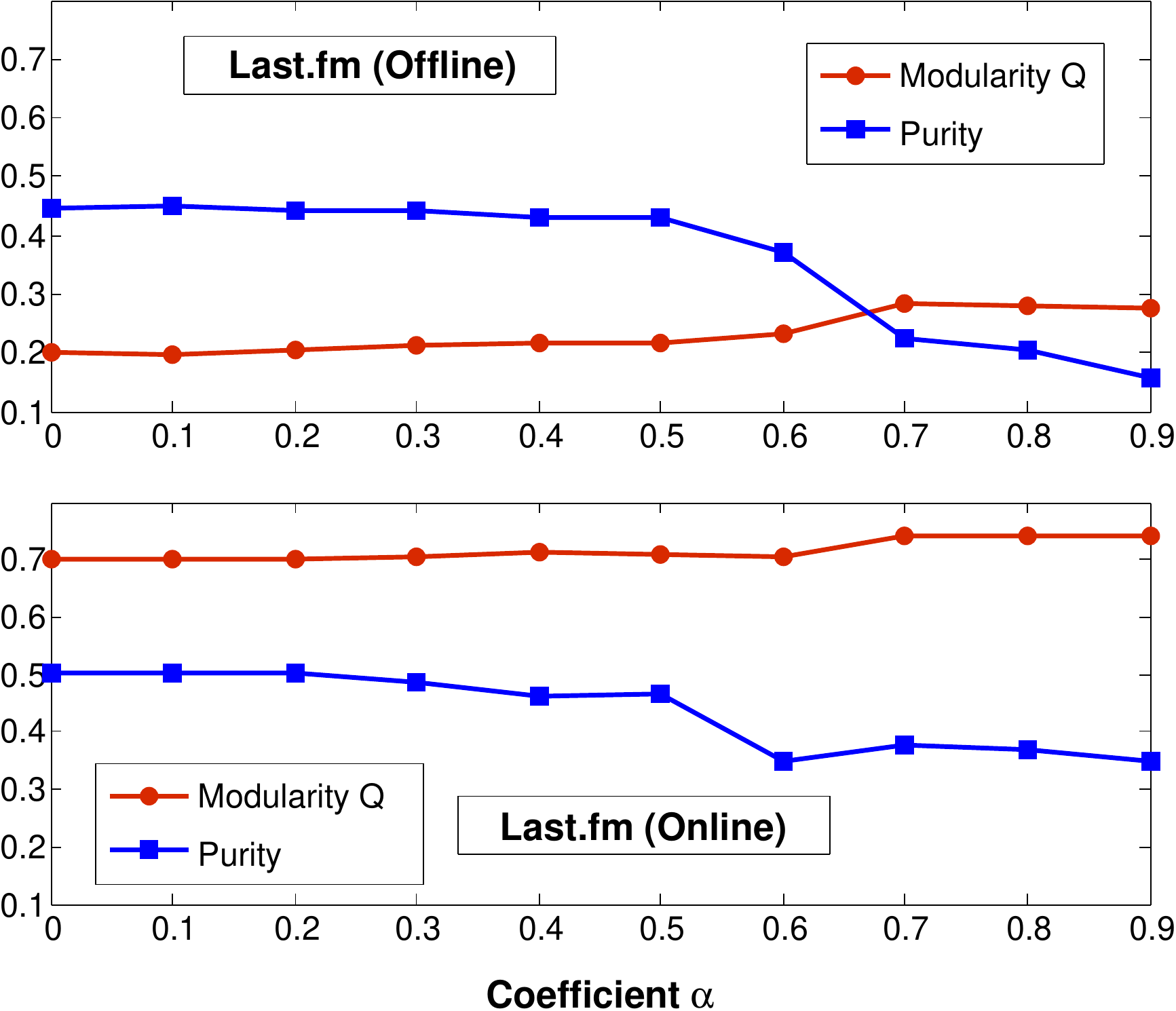}
  \caption{The evolution of Q and Purity with $\alpha$}
  \label{fig:alpha-impact}
\end{figure}

We firstly evaluate how the coefficient $\alpha$ (used in Equation~\ref{eq:event-sim}) affects the performance of our approach. Figure~\ref{fig:alpha-impact} shows the evolution of the $Purity$ and the modularity $Q$ when $\alpha$ increases. It is clear that more we put emphasis on event similarity in the latent user space, more the modularity increases. However, these metrics do not evolve at the same scale. We can observe that when the modularity slightly increases, the $Purity$ drastically decreases. Thus, good $PurQ_\beta$ score can be obtained when $\alpha \in $[0.1,0.5]. 

Then, we compare our approach with some state-of-art methods, namely:(1) Edge co-clustering inspired by the approach applied on location-based social network in~\cite{Wang:14}. For this approach, we consider as features the user similarity in the latent event space and the event similarity in the latent semantic space. Based on these features, Edge co-clustering uses k-means to cluster similar ``user-event'' edges. This method has been evaluated only on two datasets as it requires a very large computation time;(2) ECODE algorithm which introduces content-based virtual links in the graph and clusters similar events sharing high physical and virtual links; (3) The popular Modularity maximization method; (4) The EWKM-based method described in Section~\ref{sec:related-work}.

\begin{figure}[H]
  \centering
  \includegraphics[scale=0.35]{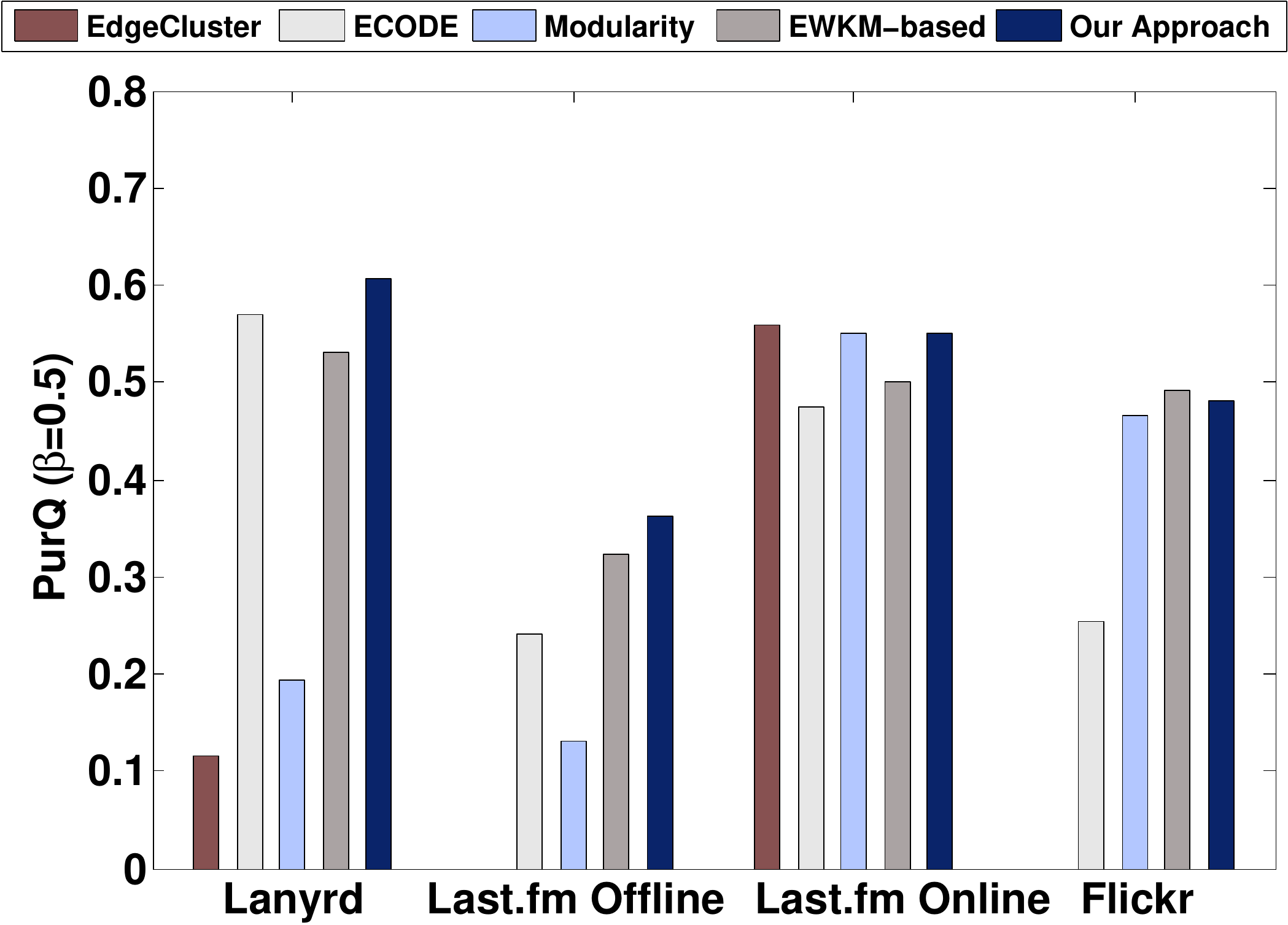}
   \includegraphics[scale=0.35]{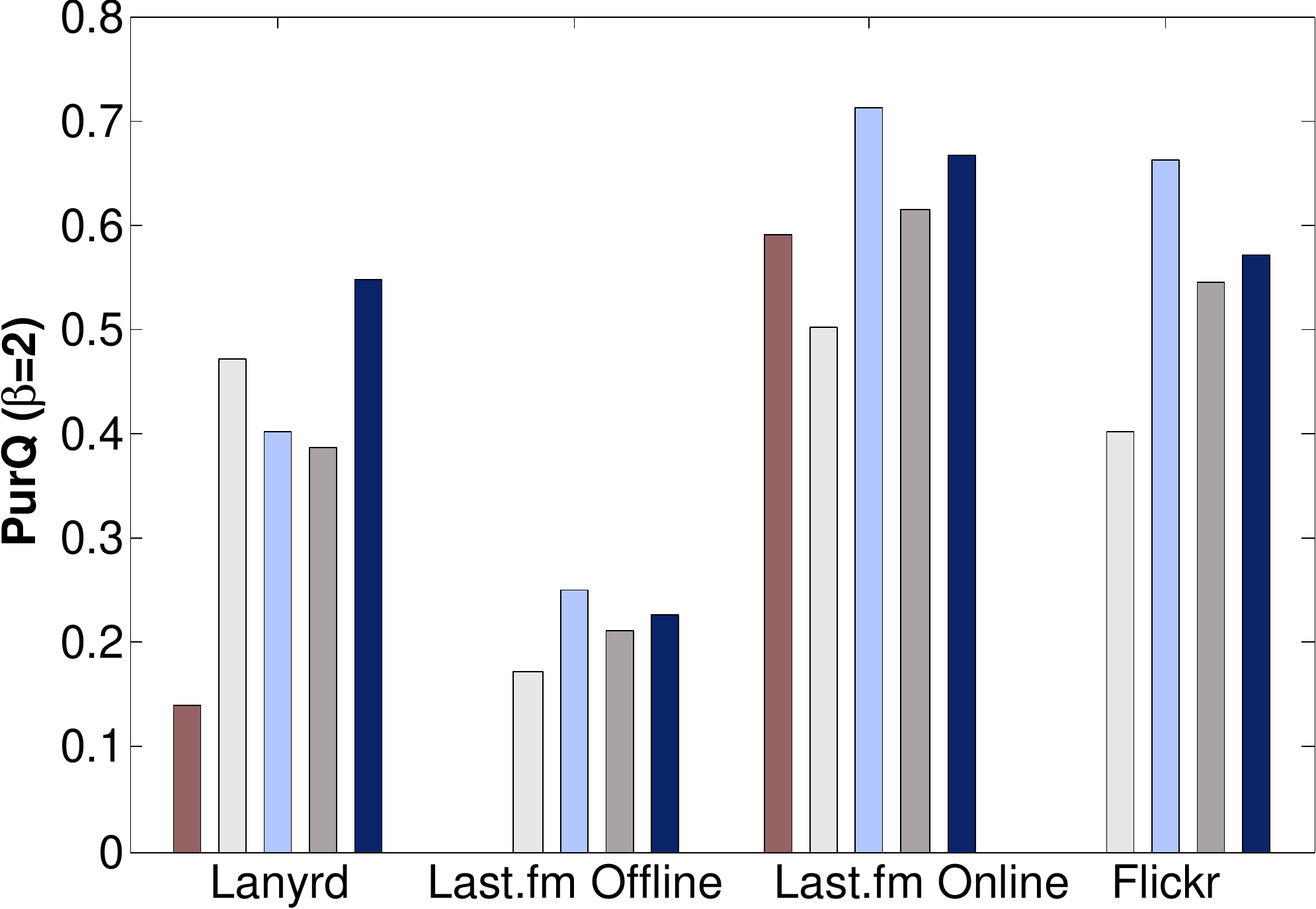}
  \caption{The performance comparison with $\beta=0.1$ and $\beta=2$ for different datasets}
  \label{fig:PurQ}
\end{figure}

The comparison results are depicted in Figure~\ref{fig:PurQ}. We observe that all the methods have nearly similar performance in Last.fm Online  ESBN particularly when $\beta=0.5$. Indeed, the communities detected have very small sizes (e.g average size=15 for the Modularity method) due to the extremely sporadic interactions. This is also explained by the low density link and the user participation behavior where 92\% of users are associated at most with only two events. Hence, the link information was sufficient to obtain a good purity. This aspect slightly decreases in Flickr dataset where 78\% of users are associated with at most two events. The modularity method apparently achieves a good purity. However, the fraction $F_{purity}$ is only equal to 0.6, a fair value compared with EWKM-based method and our approach where $F_{purity}$ are respectively equal to 0.89 and 0.91. In Last.fm Offline and Lanyrd ESBNs, the link-based method has a poor performance when $\beta=0.5$ which is explained by the higher density than the other datasets. Moreover, the identified communities are very large. For example, we found an average size of 474.5 in the communities produced by the modularity method in Last.fm offline ESBN. This indicates that the users of this network are densely linked which also explain the low $Q$ values produced by the different approaches. 
 
Comparing the content-based methods, we note a better performance for ECODE  in Lanyrd dataset than in the other datasets. This is due to the addition of virtual links to the graph based on the content-similarity between users.  However, the user profile in Last.fm is much more topically diverse than in Lanyrd which leads to ambiguous similar scores. In reality, the user may be interested in many musical concerts having different topics, whereas he has more restrictive ``scientific'' interests that mostly fit his expertise domain. From the results, we also observe a poor performance of the edge-centric clustering algorithm in Lanyrd since there is no objective function and it is sensitive to the number of cluster that need to be specified. Finally, our approach achieves the best performance both when $\beta=0.5$ and $\beta=2$. Note that there is a similar behavior between our method and the EWKM-based method.  For instance, the average size of communities in Last.fm Offline ESBN is equal to 0.33 for EWKM-based, and 0.29 for our approach. However, the EWKM-based method is based on k-means clustering which is sensitive to the initial distribution of centroids producing different results in each run. This problem is omitted in our approach since we use a hierarchical clustering. From the computation point of view, we observe that these methods have nearly the same computational time except of  the edge clustering. Finally, we also note the low purity values in Last.fm Offline ESBN compared with Lanyrd ESBN. The reason of this lower performance is that the musical concerts are attached to much more topically diverse tags than the conferences in Lanyrd. In the rest of this paper, we select the EWKM-based method to further evaluate our approach.

\begin{figure}[H]
  \centering
  \includegraphics[scale=0.37]{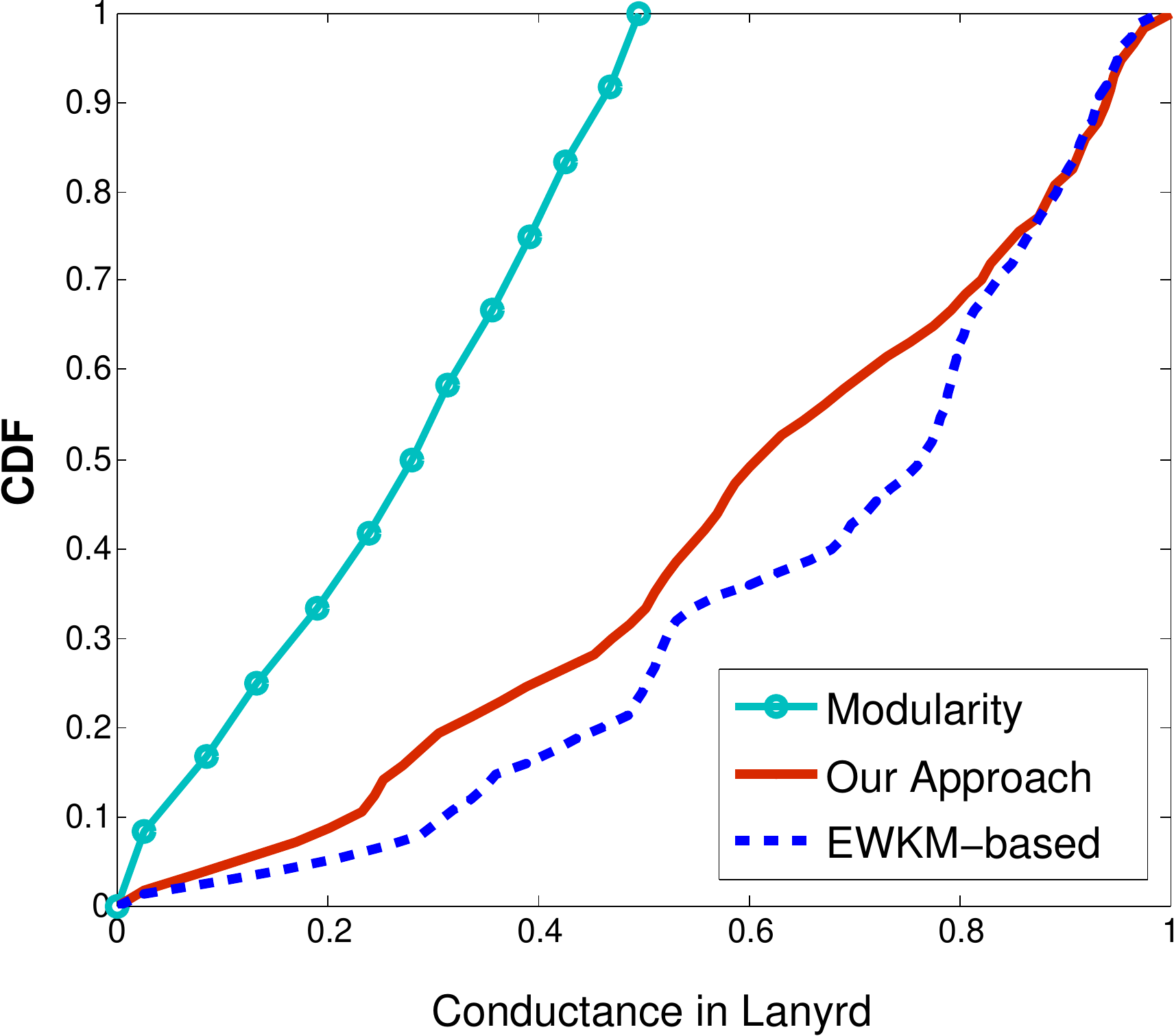}
  \caption{Conductance comparison in Lanyrd ESBN}
  \label{fig:conductance1}
\end{figure}

\begin{figure}[H]
  \centering
  \includegraphics[scale=0.37]{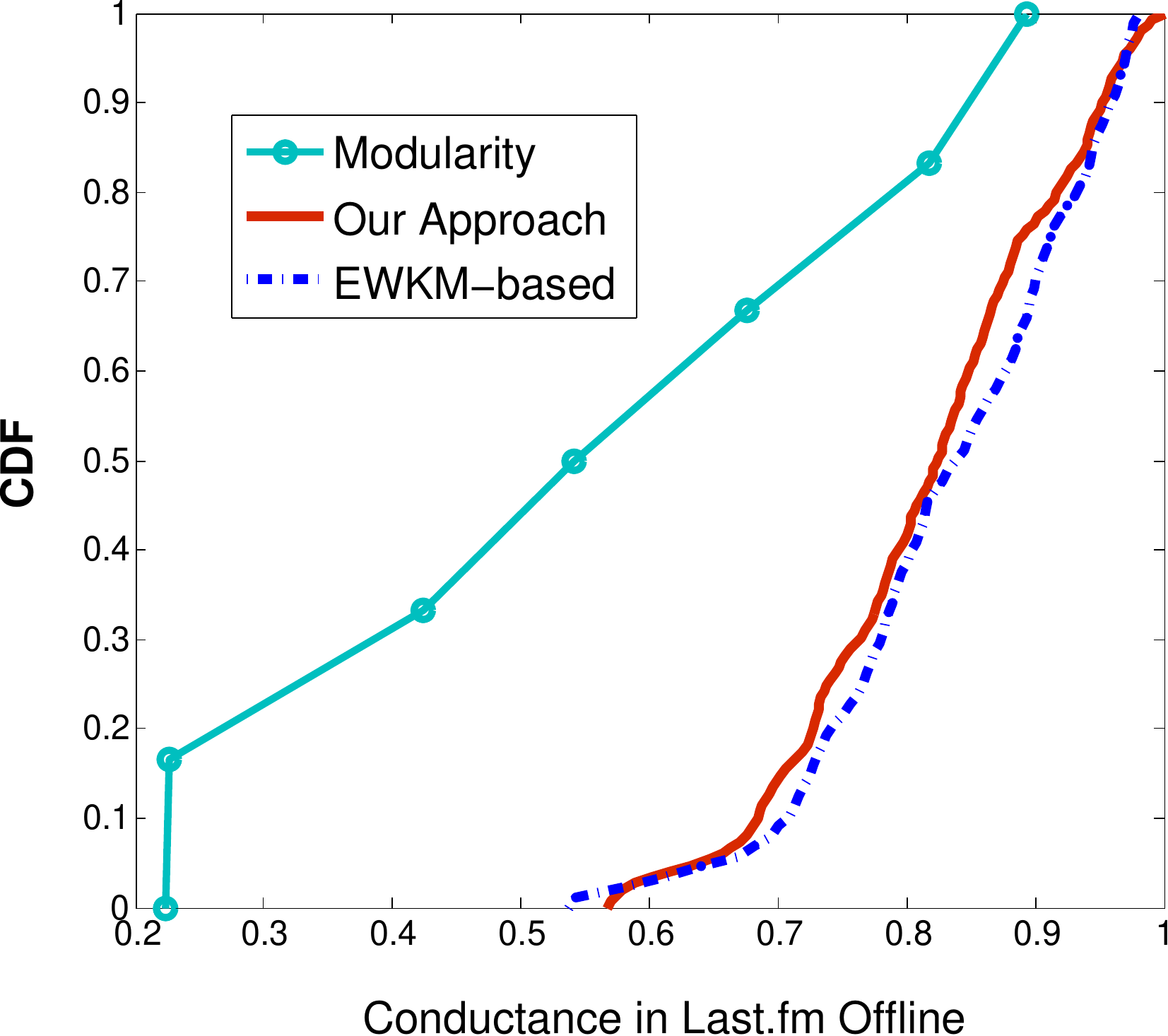}
  \caption{Conductance comparison in Last.fm Offline ESBN}
  \label{fig:conductance2}
\end{figure}

\vspace{3mm}
\textbf{Coductance Comparison} 
Since we do not have a ground truth about the real communities, we attempt to assess the proposed approach using the \emph{Conductance} metric~\cite{Leskovec:WWW08}, a popular quality function measuring if the detected communities are densely linked and attached to the rest of the
network via few edges. Note that this metric will evaluate our method from the link perspective. Lower conductance values means better community structure. Figures~\ref{fig:conductance1} and~\ref{fig:conductance2} show the cumulative distribution of the conductance metric respectively in Lanyrd and Last.fm Offline ESBNs. We can see that our approach produces slightly more communities with lower conductance values especially in Lanyrd ESBN. The reason behind these results is our strategy of the user assignment based on the link information. We believe that the better performance in Lanyrd is explained by its clustering coefficient which is larger than that of Last.fm Offline ESBN.

\begin{figure}[H]
  \centering
  \includegraphics[scale=0.44]{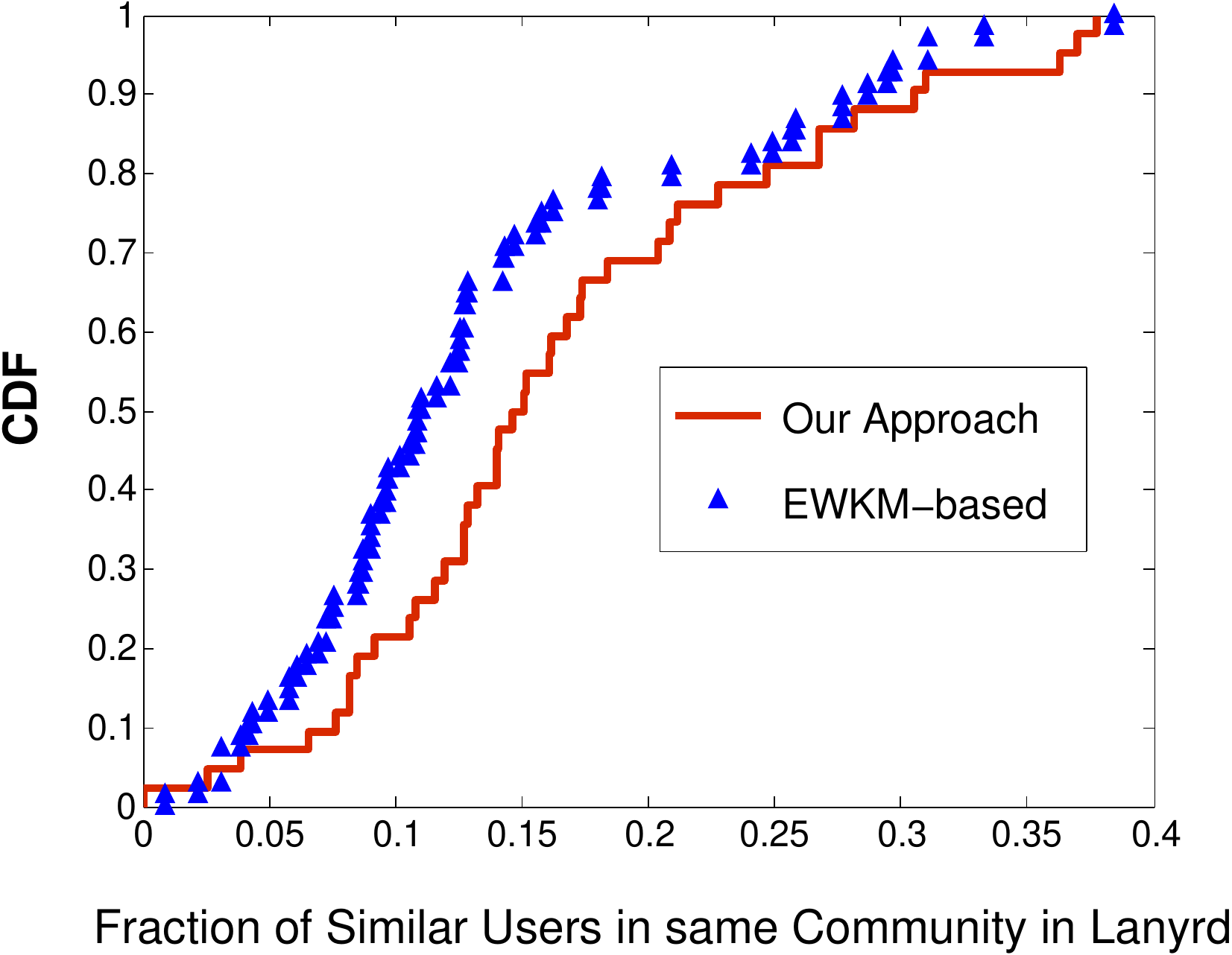}
  \caption{User Profile Comparison in Lanyrd ESBN}
  \label{fig:comp-user-1}
\end{figure}

\begin{figure}[H]
  \centering
  \includegraphics[scale=0.44]{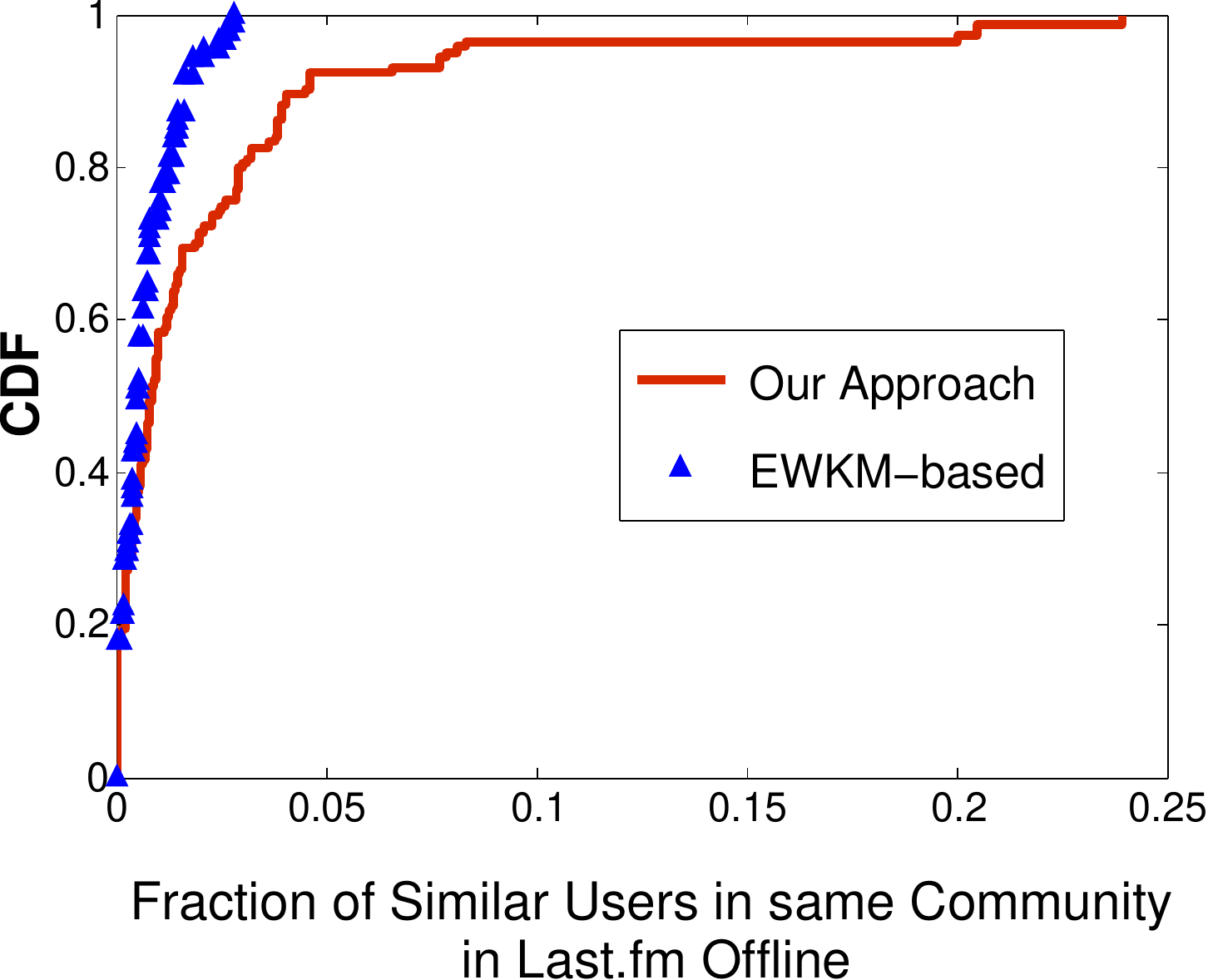}
  \caption{User Profile Comparison in Last.fm Offline ESBN}
  \label{fig:comp-user-2}
\end{figure}

\vspace{3mm}
\textbf{User Profile Comparison} 
To evaluate the methods from the content perspective, one way is to compare the user profiles within one community. Hence, we retrieved the users' tags from each website and we only keep the frequent ones. Cosine distance is then applied to compute the similarity between users. We consider that two users are similar when they have a Cosine distance above 0.3, a quite reasonable value compared with the noise of tags. Figures~\ref{fig:comp-user-1} and~\ref{fig:comp-user-2} show the cumulative distribution of the fraction of similar users within communities. We observe higher fraction values indicating that our approach groups together more topically similar users than EWKM-based method does. 

We also investigate the fraction of ``friends''  within each community. The friend relationships has been retrieved using the online social networks that exist in Last.fm and Twitter. Results are shown in Table~\ref{tab:friends}. We can see that a large fraction of friends are placed in the same community by the Modularity method in Last.fm Offline ESBN compared with the other methods. This is also justified by the very high average size of communities detected which is equal to 474.5. Moreover, it is clearly shown that the conference attendees having similar topical interests are more likely to be friends than the concert attendees, which also fits the reality. 

\begin{table}[H]
\centering{
\begin{tabular}{ccc}
\hline
    \textbf{Method}    &   Lanyrd (Twitter) & Last.fm Offline \\
\hline
   Modularity-based  &  0.72 & 0.69\\
\hline
   EWKM-based  &   0.70 &0.23\\
\hline
   Our Approach  & 0.73   & 0.29\\
\hline
\end{tabular}
\caption{Average fraction of friends within communities}
\label{tab:friends}
}
\end{table}

\textbf{Communities Overlap} 

\begin{figure}[H]
  \centering
  \includegraphics[width=\linewidth]{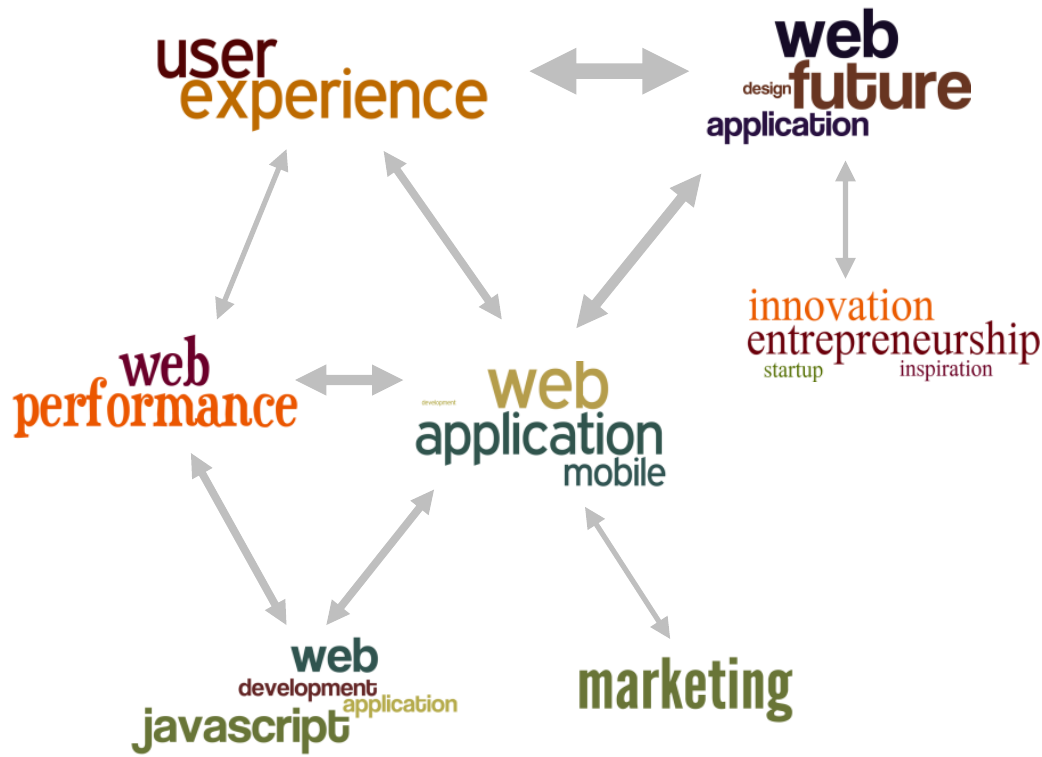}
  \caption{A sample of some overlapping communities in Lanyrd}
  \label{fig:tagcloud}
\end{figure}

Finally, Figure~\ref{fig:tagcloud} shows a tag cloud representing a sample of the most overlapping communities in Lanyrd ESBN. We only represent the links that exhibit the high overlapping degree. We can understand that the main topic of these communities is the web domain which is the interest of many users of different ``topical'' expertise. 

In Lanyrd ESBN, our approach detects 65 communities while the EWKM-based produces 92 communities. Analyzing both communities, it is found that we discover fewer but more cohesive topical communities. Note that we evaluate the cohesiveness using the popular Silhouette coefficient~\cite{Rousseeuw:1987}. For instance, we detect only one community about \emph{user experience} with a cohesion equal to 0.1.
While the EWKM-based method detects 4 communities containing 2 ``singletons'' about \emph{user experience} and having a cohesion equal to -0.3. We also justify this higher cohesion by our strategy to assign users bringing together strongly linked users.

%
%

\section{Conclusion}
\label{sec:conclusion}
Today's people are massively using the event-based services to interact together, either online by sharing comments and photos or offline by attending events. Moreover, the social connections can be formed and strengthened during social events which can be considered as a mean to detect communities. In this paper, we have proposed a new hierarchical based approach to mine topical communities from event information. Taking into account both the content and the link information, we first perform the event clustering by maximizing a new defined metric called \emph{Semantic Modularity}. Then all participants involved in those events are partitioned by evaluating an attachment score based on the user's degree. Extensive experimental results have shown the efficiency of our approach to ensure high purity and modularity measures, compared with existing methods.

For future work, we plan to combine the offline and online ESBNs as it is considered as an heterogeneous network. We would like to evaluate the impact of this combination on the purity and the modularity of topical clusters.

\bibliographystyle{abbrv}
\bibliography{kdd14}

\begin{thebibliography}{10}

\bibitem{Blei:MLR03}
D.~M. Blei, A.~Y. Ng, and M.~I. Jordan.
\newblock Latent dirichlet allocation.
\newblock {\em Journal of Machine Learning Research}, 3:993--1022, 2003.

\bibitem{Clauset:PR04}
A.~Clauset, M.~E.~J. Newman, and C.~Moore.
\newblock Finding community structure in very large networks.
\newblock {\em Physical Review E}, 70:066111, 2004.

\bibitem{Juan:cason11}
J.~D. Cruz, C.~Bothorel, and F.~Poulet.
\newblock Entropy based community detection in augmented social networks.
\newblock In {\em International Conference on Computational Aspects of Social
  Networks (CASoN)}, pages 163--168, Salamanca, Spain, 2011.

\bibitem{Dhillon:KDD01}
I.~S. Dhillon.
\newblock Co-clustering documents and words using bipartite spectral graph
  partitioning.
\newblock In {\em 7$^{th}$ ACM SIGKDD International Conference on Knowledge
  Discovery and Data Mining}, New York, NY, USA, 2001.

\bibitem{Griffiths:04}
T.~L. Griffiths and M.~Steyvers.
\newblock {Finding scientific topics}.
\newblock {\em National Academy of Sciences of the United States of America},
  101:5228--5235, 2004.

\bibitem{Han:NCWTW12}
J.~Han, J.~Niu, A.~Chin, W.~Wang, C.~Tong, and X.~Wang.
\newblock How online social network affects offline events: A case study on
  douban.
\newblock In {\em 9$^{th}$ International Conference on Ubiquitous Intelligence
  and Computing}, Fukuoka, September, 2012.

\bibitem{Thomas:99}
T.~Hofmann.
\newblock Probabilistic latent semantic indexing.
\newblock In {\em 22$^{nd}$ Annual International ACM SIGIR Conference on
  Research and Development in Information Retrieval}, pages 50--57, Berkeley,
  CA, USA, 1999.

\bibitem{Khrouf:RAMSS12}
H.~Khrouf, G.~Atemezing, G.~Rizzo, R.~Troncy, and T.~Steiner.
\newblock {Aggregating Social Media for Enhancing Conference Experience}.
\newblock In {\em 1$^{st}$ International Workshop on Real-Time Analysis and
  Mining of Social Streams (RAMSS'12)}, Dublin, Ireland, 2012.

\bibitem{Khrouf:ISWC12}
H.~Khrouf, V.~Milicic, and R.~Troncy.
\newblock Eventmedia live: Exploring events connections in real-time to enhance
  content.
\newblock In {\em Semantic Web Challenge at 11$^{th}$ International Semantic
  Web Conference}, Boston, USA, 2012.

\bibitem{Khrouf:SWJ12}
H.~Khrouf and R.~Troncy.
\newblock Eventmedia: a {LOD} dataset of events illustrated with media.
\newblock {\em Semantic Web Journal, {S}pecial {I}ssue on {L}inked {D}ataset
  descriptions}, 2012.

\bibitem{Leskovec:WWW08}
J.~Leskovec, K.~J. Lang, A.~Dasgupta, and M.~W. Mahoney.
\newblock Statistical properties of community structure in large social and
  information networks.
\newblock In {\em 17$^{th}$ International Conference on World Wide Web}, WWW
  '08, pages 695--704, New York, NY, USA, 2008.

\bibitem{Li:2013}
L.~Li and N.~Memon.
\newblock Mining groups of common interest: Discovering topical communities
  with network flows.
\newblock In {\em 9$^{th}$ International Conference on Machine Learning and
  Data Mining in Pattern Recognition}, Berlin, Heidelberg, 2013.

\bibitem{Li:dasfa11a}
X.~Li, A.~Tan, P.~S. Yu, and S.-K. Ng.
\newblock Ecode: Event-based community detection from social networks.
\newblock In {\em Database Systems for Advanced Applications}, pages 22--37,
  2011.

\bibitem{Liu:KDD12}
X.~Liu, Q.~He, Y.~Tian, W.-C. Lee, J.~McPherson, and J.~Han.
\newblock Event-based social networks: Linking the online and offline social
  worlds.
\newblock In {\em 18$^{th}$ ACM SIGKDD conference on Knowledge Discovery and
  Data Mining}, Beijing, China, 2012.

\bibitem{McPherson:2001}
M.~McPherson, L.~Smith-Lovin, and J.~M. Cook.
\newblock Birds of a feather: Homophily in social networks.
\newblock {\em Annual Review of Sociology}, 27(1):415--444, 2001.

\bibitem{Newman:04}
M.~E.~J. Newman and M.~Girvan.
\newblock Finding and evaluating community structure in networks.
\newblock {\em Phys. Rev. E}, 69:026113, 2004.

\bibitem{Rousseeuw:1987}
P.~Rousseeuw.
\newblock Silhouettes: A graphical aid to the interpretation and validation of
  cluster analysis.
\newblock {\em J. Comput. Appl. Math.}, 20(1):53--65, 1987.

\bibitem{Sachan:WWW12}
M.~Sachan, D.~Contractor, T.~A. Faruquie, and L.~V. Subramaniam.
\newblock Using content and interactions for discovering communities in social
  networks.
\newblock In {\em 21$^{st}$ World Wide Web Conference}, Lyon, France, 2012.

\bibitem{Shaghayegh:RSWEB11}
S.~Sahebi and W.~Cohen.
\newblock Community-based recommendations: a solution to the cold start
  problem.
\newblock In {\em 3$^{rd}$ Workshop on Recommender Systems and the Social Web
  (RSWEB)}, Chicago, IL, USA, 2011.

\bibitem{Shaw:ASWC09}
R.~Shaw, R.~Troncy, and L.~Hardman.
\newblock {LODE: Linking Open Descriptions Of Events}.
\newblock In {\em 4$^{th}$ Asian Semantic Web Conference (ASWC'09)}, pages
  153--167, Shanghai, China, 2009.

\bibitem{Steyvers:kdd04}
M.~Steyvers, P.~Smyth, M.~Rosen-Zvi, and T.~L. Griffiths.
\newblock Probabilistic author-topic models for information discovery.
\newblock In {\em ACM SIGKDD International Conference on Knowledge Discovery
  and Data Mining}, pages 306--315, 2004.

\bibitem{Wang:14}
Z.~Wang, X.~Zhou, D.~Zhang, D.~Yang, and Z.~Yu.
\newblock Cross-domain community detection in heterogeneous social networks.
\newblock {\em Personal and Ubiquitous Computing}, 18(2):369--383, 2014.

\bibitem{Scott:05}
S.~White and P.~Smyth.
\newblock A spectral clustering approach to finding communities in graph.
\newblock In {\em SIAM International Conference on Data Mining}, pages
  274--285, Newport Beach, CA, USA, 2005.

\bibitem{Zhongying:12}
Z.~Zhao, S.~Feng, Q.~Wang, J.~Z. Huang, G.~J. Williams, and J.~Fan.
\newblock Topic oriented community detection through social objects and link
  analysis in social networks.
\newblock {\em Knowledge-Based Systems}, 26:164--173, 2012.

\bibitem{Zhou:WWW06}
D.~Zhou, E.~Manavoglu, J.~Li, C.~L. Giles, and H.~Zha.
\newblock Probabilistic models for discovering e-communities.
\newblock In {\em 15$^{th}$ International Conference on World Wide Web}, pages
  173--182, New York, NY, USA, 2006.

\end{thebibliography}

\end{document}